\documentclass[11pt,twoside]{article}

\usepackage{asp2004}
\usepackage{epsf}
\usepackage{psfig}
\usepackage{lscape}

\begin{document}
\title{The nature of (sub)millimetre galaxies in hierarchical models}  

\author{
C. M. Baugh$^{1}$,
C. G. Lacey$^{1}$,
C. S. Frenk$^{1}$,
G. L. Granato$^{2}$,
L. Silva$^{3}$,
A. Bressan$^{2}$,
A. J. Benson$^{4}$,
S. Cole$^{1}$.}   
\affil{
$^{1}$Institute for Computational Cosmology, Department of Physics, 
University of Durham, South Road, Durham DH1 3LE, UK. \\
$^{2}$Osservatorio Astronomico di Padova, Vicolo dell'Osservatorio, 5, I-35122 Padova, Italy.\\
$^{3}$Osservatorio Astronomico di Trieste, via Tiepolo 11, I34131 Trieste, Italy.\\
$^{4}$Astrophysics, University of Oxford, Keble Road, Oxford, OX1 3RH, UK. 
}

\begin{abstract} 
We present a hierarchical galaxy formation model which can account 
for the number counts of sources detected through their emission at 
sub-millimetre wavelengths. 
The first stage in our approach is an {\it ab initio} calculation of 
the star formation histories for a representative sample of galaxies,  
which is carried out using the semi-analytical galaxy 
formation model {\tt GALFORM}. 
These star formation histories are then input into the 
spectro-photometric code {\tt GRASIL}, to produce a spectral energy 
distribution for each galaxy. 
Dust extinction and emission are treated self consistently in our model, 
without having to resort to ad-hoc assumptions about the amount of 
attenuation by dust or the temperature at which the dust radiates. 
We argue that it is necessary to modify the form of the stellar initial 
mass function in starbursts in order to match the observed number of 
sub-mm sources, if we are to retain the previous good matches enjoyed 
between observations and model predictions in the local universe. We also 
list some other observational tests that have been passed by our model.  
\end{abstract}

\section{Introduction}

\begin{figure}
{\epsfxsize=10.truecm
\epsfbox[-100 150 590 720]{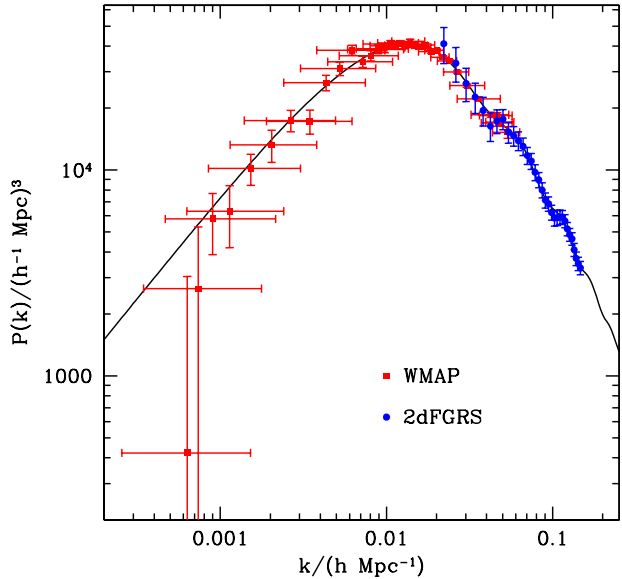}}
\caption{ 
The power spectrum of temperature fluctuations in the cosmic microwave 
background radiation (the first year WMAP data; Hinshaw et~al. 2003) 
plotted on the same scale as the power spectrum of galaxy clustering 
measured from the two-degree field galaxy redshift survey by 
Cole et~al. (2005). Figure courtesy of Ariel Sanchez, based on results 
from Sanchez et~al. (2006). 
}
\label{fig:pk}
\end{figure}

The first few years of the millennium have seen increased support assembled 
for the hierarchical structure formation paradigm (Spergel et~al. 2003). 
In this model, small ripples in the density of the primordial universe 
are amplified by the force of gravity acting over billions of years. 
The most successful model, a universe in which cold dark matter outweighs 
baryonic matter and in which the rate of expansion is accelerating due 
to a dynamically dominant dark energy component, agrees spectacularly 
well with the latest measurements of the primordial spectrum of density 
perturbations, as shown in Fig.~\ref{fig:pk}. 

Physical models of galaxy formation in a hierarchical universe are also 
reaching maturity. The roots of modern-day ``semi-analytical'' galaxy 
formation models actually predate the cold dark matter cosmology and  
took hold in the 1970s, with the papers by Press \& Schechter (1974) 
and White \& Rees (1978). These papers set out the basic ideas which underpin 
the approach, namely that galaxies form by the radiative cooling of baryons 
inside dark matter haloes which were assembled by a merging process driven 
by gravitational instability. The 1990s saw the first detailed calculations 
based on these ideas which established the validity 
of the approach (White \& Frenk 1991; Kauffmann, White \& Charlot 1993; 
Lacey et~al. 1993; Cole et~al. 1994). These models are now firmly 
established as a powerful tool which can generate testable predictions, 
connecting hierarchical clustering cosmologies to observations of the 
galaxy population at different epochs in the history of the universe. 

Now that the parameter space which defines the background cosmology is 
shrinking (see for example Sanchez et~al. 2006), semi-analytical models 
of galaxy formation are entering a new phase. Coupled with the explosion 
of observational data available for galaxies at high redshift, the focus 
is shifting towards a critical assessment of the physics implemented in 
the models. The modular nature of the models and their speed means that 
different prescriptions can be tested for a particular physical process. 
The ongoing efforts to improve the modelling of the various phenomena 
involved in the galaxy formation process are inevitable, given their 
complexity and our lack of detailed understanding of the relevant physics.

\begin{figure}[t]
{\epsfxsize=12.truecm
\epsfbox[-50 150 590 690]{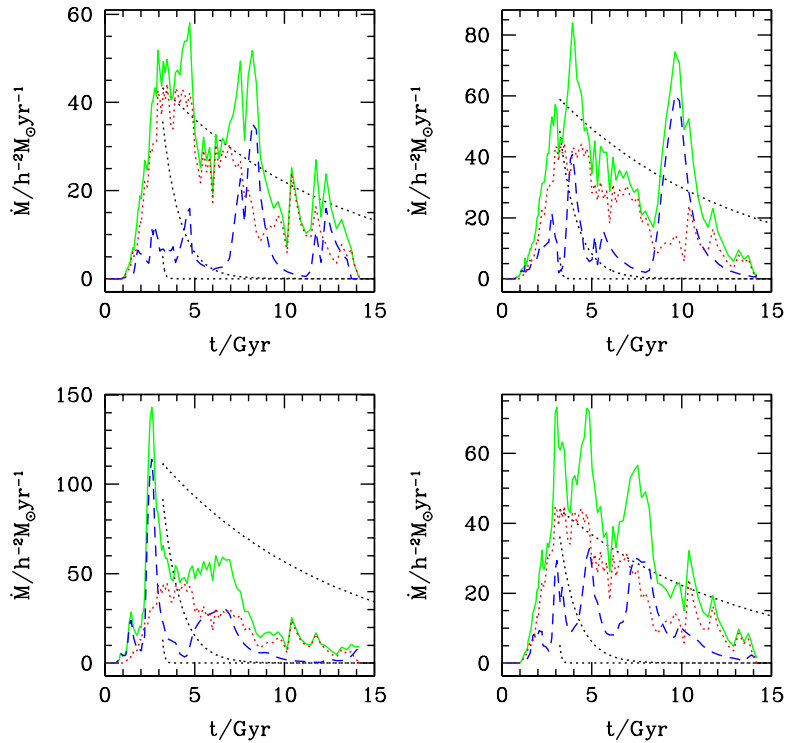}}
\caption{ 
The star formation histories for four massive 
galaxies, as predicted by the {\tt GALFORM} model, 
plotted as a function of the age of the model universe. 
The total star formation rate (solid line) takes into account 
the quiescent star formation in the progenitors of the present 
day object (dashed line), and includes starbursts triggered by 
galaxy mergers (dotted line). The smooth dotted curves show 
simple exponential star formation laws for comparison; these 
star formation histories start when the universe was 3\,Gyr old 
and have e-fold times of 0.1, 1 and 10 Gyr.  
}
\label{fig:sfhist}
\end{figure}

The problem of matching the bright end of the local field galaxy luminosity 
function provides a good illustration of this point (Benson et~al. 2003). 
The first generation of semi-analytical models had little problem in 
matching the exponential break observed in galaxy luminosity function. 
Today, modellers find this a more challenging task for two reasons: 
(i) A shift in the favoured cosmological model. Today, the ``standard'' 
cold dark matter model has a matter density parameter less than one third of  
the critical density (Sanchez et~al. 2006). 
Structures tend to form earlier in a dark energy dominated universe, 
so, coupled with the slightly older age for the universe in this 
cosmology, more massive haloes have been able to cool gas for a longer 
period than would have been the case in a universe with the critical 
density in matter. This leads to more gas cooling in these haloes. This 
problem is exacerbated by the tighter observational constraints on the 
baryon density, which typically result in higher baryon densities being 
input into the models than would have been used in earlier calculations. 
(ii) The luminosity function is not being considered in isolation. The 
continued development of the models now means that they are able to predict 
a much wider range of galaxy properties than was possible in the early days. 
This increased sophistication actually makes it more difficult to match 
one particular observation because other galaxy properties can be 
adversely affected by parameter changes. 

The attempt to find a solution to the problem of matching the sharpness 
of the observed break of the bright end of the luminosity function 
has led to a revision of the treatments of gas cooling and feedback 
in massive haloes used in the models 
(e.g. Benson et~al. 2003; Bower et~al. 2006; Croton et~al. 2006; 
de Lucia et~al. 2006). While some 
of our more conservative colleagues view such changes as sufficient  
grounds on which to dismiss the semi-analytical approach altogether, 
what we are witnessing is simply the application of the scientific method; 
when a model prediction is found to be incorrect, this shows that 
an ingredient in the scenario is 
either modelled incorrectly or is missing altogether. The resolution of 
the discrepancy leads to a new model in which our understanding of 
galaxy formation has been advanced. Now that the utility of this approach 
has gained general acceptance, we should welcome conflict between 
observations and theoretical predictions, as this will drive future 
progress in the models. 

In this article, we deal with another area in which the models have faced 
a stern challenge; matching the abundance of high redshift galaxies detected 
through their dust emission in the sub-mm. At first sight, the galaxies 
seen with the {\tt SCUBA} instrument at 850 microns appeared to be 
massive galaxies at high redshift, with star formation rates approaching 
$1000 {\rm M}_{\odot} {\rm yr}^{-1}$ 
(Smail et~al. 1997). Such objects would dominate the star formation in the 
early Universe, dwarfing the contribution of galaxies seen in the rest-frame 
ultraviolet (Hughes et~al. 1998; Barger et~al. 1998). Our solution to this 
problem is controversial, but spawns a number of testable predictions. 
On the whole these predictions agree remarkably well with observations, 
as we will discuss. In \S~2, we give a brief overview 
of our model of galaxy formation. Our treatment of the 
impact of dust on the spectral energy distribution  
of our model galaxies is a novel apsect of our model and is described 
in \S~3. We present the main results of interest to 
the participants of this workshop in \S~4. 
Further tests of the model are listed in \S~5 along with our conclusions.

\section{The galaxy formation model}
\label{sec:model}

We use the semi-analytical model {\tt GALFORM}; the content  
of the model and the philosophy behind it are set out in detail in 
Cole et~al. (2000). Important revisions to the basic model are described 
in Benson et~al. (2002, 2003). Our solution to the problem of accommodating 
the number counts of {\tt SCUBA} sources in the cold dark matter model 
is explained in Baugh et~al. (2005).

In summary, the aim of {\tt GALFORM} is to carry out an {\it ab initio} 
calculation of the formation and evolution of galaxies, in a background 
cosmology in which structures grow hierarchically. The physical ingredients 
considered in the model include: 
(i) The formation of dark matter haloes through mergers and accretion 
of material. 
(ii) The collapse of baryons into the gravitational potential wells of 
dark matter haloes. 
(iii) The radiative cooling of gas that is shock heated during infall 
into the dark halo. 
(iv) The formation of a rotationally supported disk of cold gas. 
(v) The formation of stars from the cold gas. 
(vi) The injection of energy into the interstellar medium, through 
supernova explosions or the accretion of material onto a supermassive 
black hole. 
(vii) The chemical evolution of the interstellar medium, stars and 
the hot gas. 
(viii) The merger of galaxies following the merger of their host dark matter 
haloes, due to dynamical friction.  
(ix) The formation of spheroids during mergers due to the rearrangement of 
pre-existing stars (i.e. the disk and bulge of the progenitor 
galaxies) and the formation of stars in a burst.  
(x) The construction of a composite stellar population for each galaxy, 
yielding a spectral energy distribution, including the effects of 
dust extinction, a point to which we shall return in more detail later 
on in this section. 

Four examples of the star formation histories predicted by {\tt GALFORM} 
are shown in Fig.~\ref{fig:sfhist}. The cases shown are massive galaxies 
at the present day. The star formation history of a galaxy is constructed 
by considering the quiescent star formation in all of its progenitors and 
all the bursts of star formation triggered by galaxy mergers. A common 
{\it assumption} for the star formation history of a galaxy made in 
many other models is that stars form with an exponentially declining rate; 
some examples of such histories are marked on each panel with 
illustrative e-folding times. The star formation histories {\it predicted} 
by {\tt GALFORM} are quite different from the simple exponential form. 

\section{The effect of dust on the spectral energy distribution}
\label{sec:dust}

In order to make predictions for the sub-mm emission from galaxies, 
we need to take into account the effect of dust on the spectral 
energy distribution of galaxies. 
Previous work in this area has either employed 
template spectral energy distributions based on local galaxies (e.g. 
Blain et~al. 1999; Devriendt \& Guiderdoni 2000) or has treated the 
temperature of the dust, $T_{\rm d}$, as a free parameter (e.g. Kaviani, 
Haehnelt \& Kauffmann 2003). The dust luminosity per unit frequency at 
long wavelengths scales as $T^{-5}_{\rm d}$ for a given 
bolometric dust luminosity, for a standard assumption about the 
emissivity of the dust. 
Given this strong dependence of luminosity on $T_{\rm d}$, it would appear 
trivial to match the observed sub-mm counts by simply making a modest 
tweak to the dust temperature. Unfortunately, such a model is unphysical. 
The dust temperature should be set by requiring that the dust grains be 
in thermal equilibrium, with a balance between radiative heating 
and cooling. With this criteria met, the dust luminosity per unit frequency 
depends rather less dramatically upon the bolometric luminosity and the 
dust mass; significant changes to these properties are required to change 
the dust luminosity (see Baugh et~al. 2005 for a discussion).   

An important feature of our model is a physically consistent treatment of the 
extinction of starlight by dust and the reprocessing of this energy 
at longer wavelengths. This is achieved by using  
the {\tt GRASIL} spectro-photometric model introduced by Silva et~al. (1998). 
{\tt GRASIL} computes the emission from both the stars and dust in a galaxy, 
based on the star formation and metal enrichment histories predicted by the 
semi-analytical model (Granato et~al. 2000). {\tt GRASIL} includes 
radiative transfer through a two-phase dust medium, with a diffuse component 
and giant molecular clouds, and a distribution of dust grain sizes. 
Stars are assumed to form inside the clouds and then gradually migrate. The
output from {\tt GRASIL} is the galaxy SED from the far-UV to the sub-mm.

\section{Results}
\label{sec:res}

\begin{figure}[t]
{\epsfxsize=12.truecm
\epsfbox[-50 150 590 700]{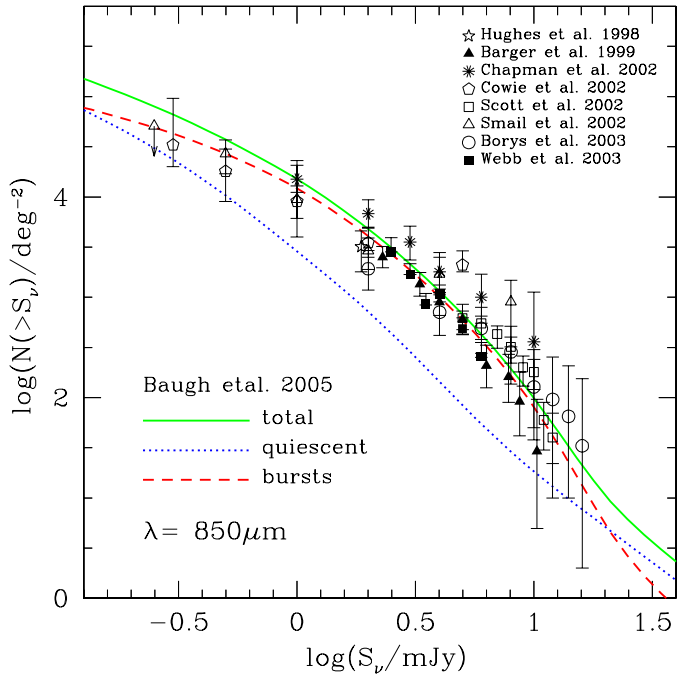}}
\caption{ 
The cumulative number counts of galaxies at 850 microns predicted 
by the Baugh et~al. (2005) model, compared with a compilation of 
observational estimates, as indicated by the legend. The solid curve 
shows the total number counts, the dashed curve the contribution from 
galaxies which are undergoing a galaxy merger induced starburst and 
the dotted curve shows the counts from galaxies that are forming stars 
quiescently in galactic discs. 
}
\label{fig:s850}
\end{figure}

\begin{figure}
{\epsfxsize=12.truecm
\epsfbox[-50 150 590 700]{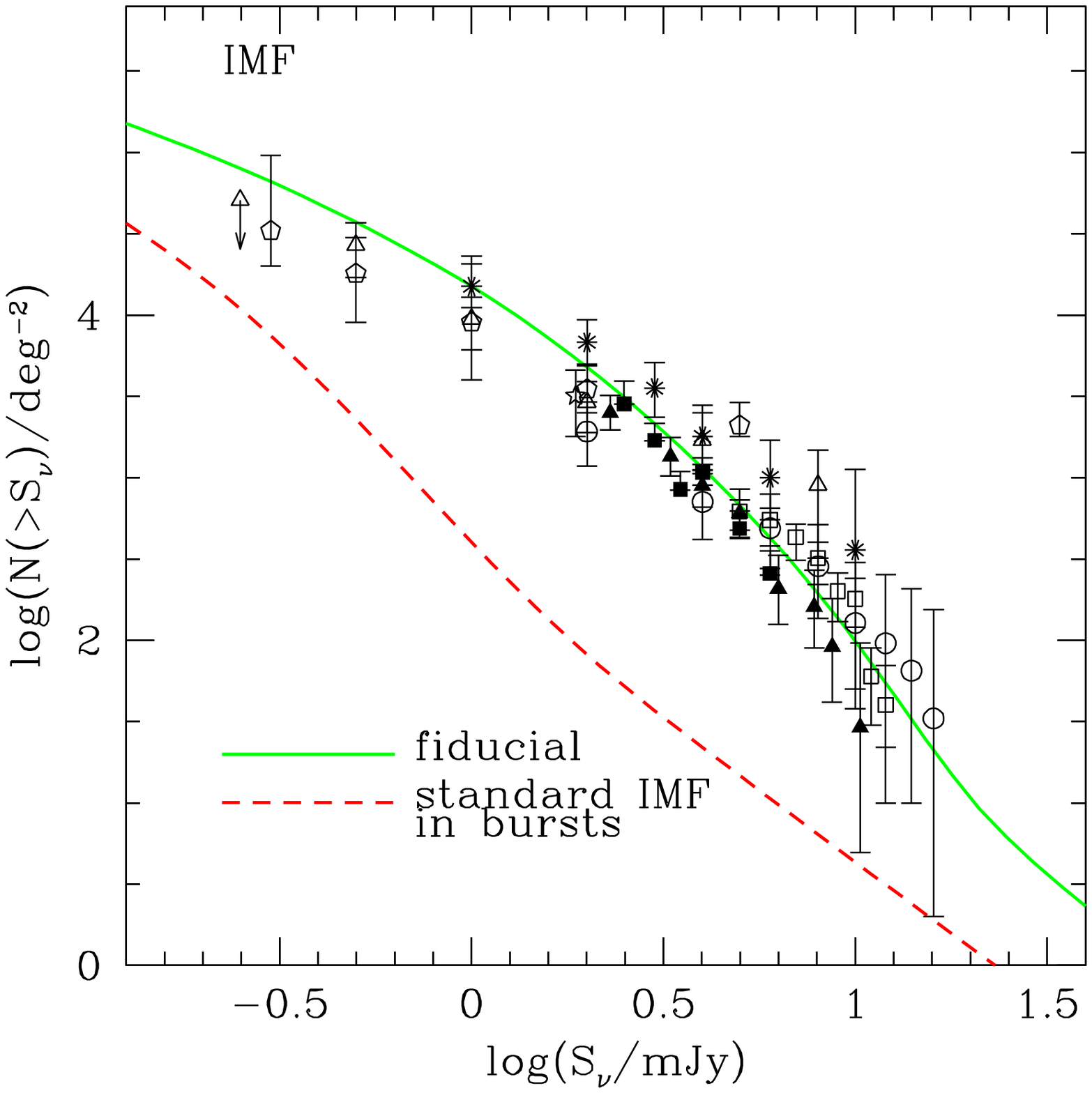}}
{\epsfxsize=12.truecm
\epsfbox[-50 150 590 700]{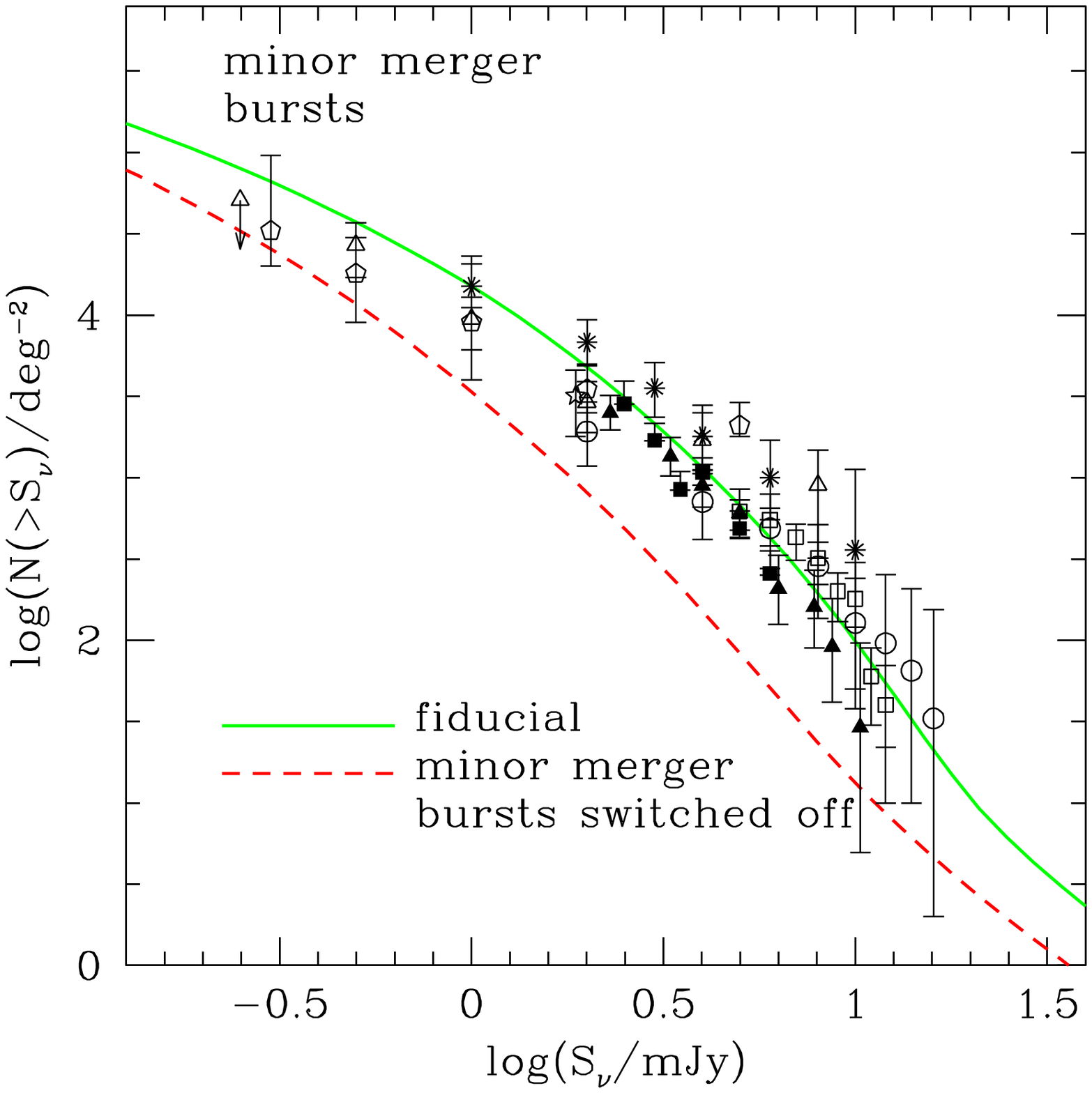}}
\caption{ 
The impact on the predicted number counts of switching off key 
ingredients of the model. The fiducial model from Baugh et~al. 
(2005) is shown by the solid line; as Fig~\ref{fig:s850} shows 
this model matches the observed counts remarkably well. 
In the upper panel, the dashed line shows the predicted counts if 
we adopt a standard IMF for star formation in merger induced bursts, 
rather than the flat IMF used in the fiducial model. 
In the bottom panel, the dashed curve shows how the counts change if 
starbursts triggered by minor merger (i.e. when a gas rich disk is hit 
by a small satellite) are switched off. 
}
\label{fig:s850comp}
\end{figure}

\begin{figure}[t]
{\epsfxsize=11.truecm
\epsfbox[-50 150 590 700]{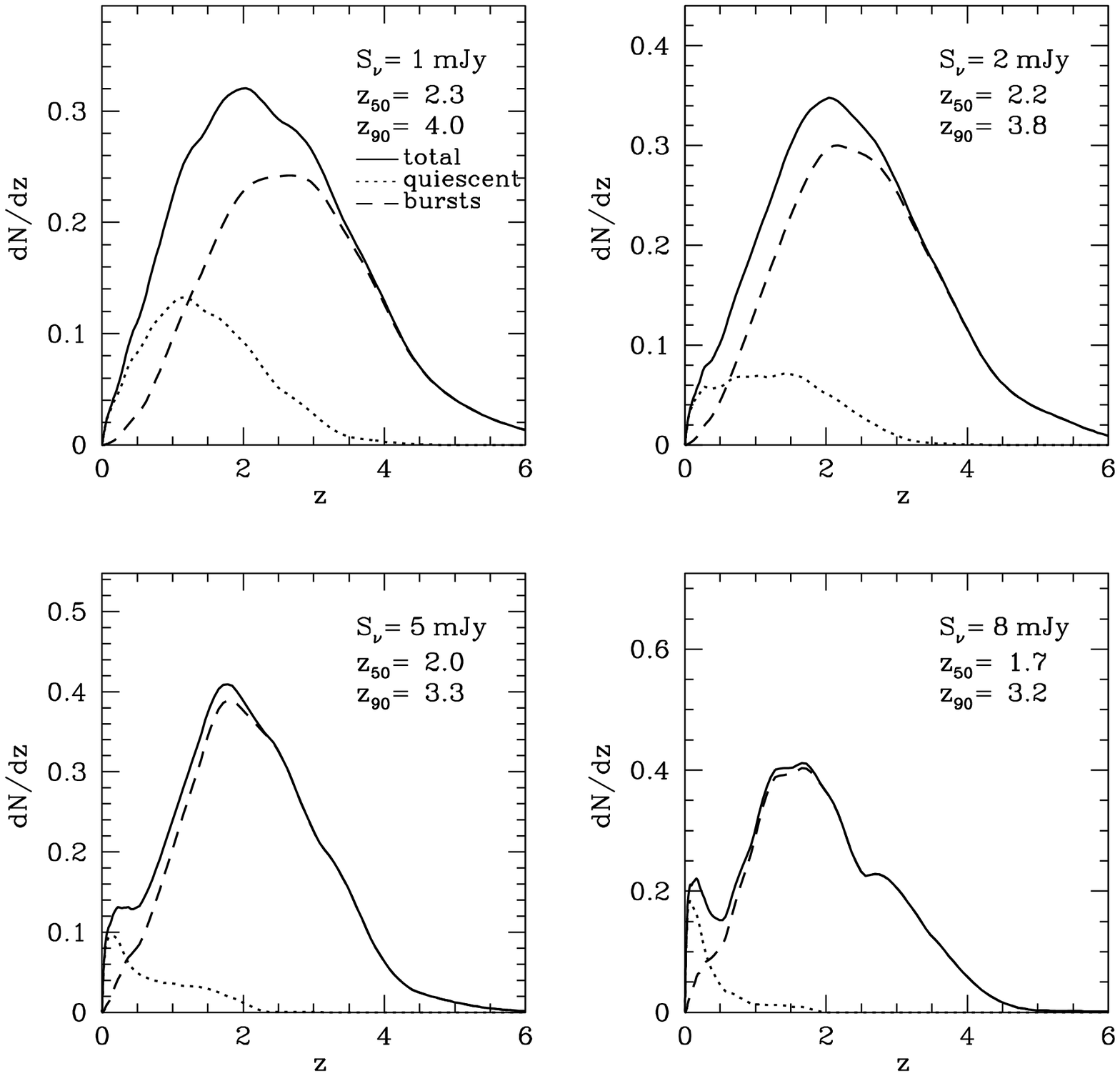}}
\caption{ 
The predicted redshift distribution at a series of 850 micron 
fluxes (as indicated in the key). The solid lines shows the 
redshift distribution of all galaxies, the dashed lines shows 
ongoing bursts and the dotted lines show galaxies which are 
forming star quiescently. The median redshift ($z_{50}$) and the 
redshift below with 90\% of galaxies are predicted to lie ($z_{90}$) are 
also given on each panel. 
}
\label{fig:dndz}
\end{figure}

Previous attempts to match the observed counts of sub-mm sources using the 
combined {\tt GALFORM} and {\tt GRASIL} machinery, whilst retaining the 
successes of the models at other redshifts, were unsuccessful, failing 
to match the counts by over an order of magnitude (see Baugh et~al. 2005). 
The are two principle reasons for the increased counts of sub-mm galaxies in 
the model introduced by Baugh et~al, as shown by Fig.~\ref{fig:s850}. 
Firstly, more star formation takes place in starbursts than in 
earlier models. There are two reasons for this. In the new model, 
the timescale for quiescent star formation is independent of redshift, 
instead of scaling with the dynamical time of the galaxy as in the fiducial 
model of Cole et~al. (2000). High redshift disks consequently have 
larger gas fractions than before, resulting in gas rich starbursts 
at early epochs. In addition, in the new model, a burst can be triggered 
by the accretion of a satellite galaxy which brings in a modest amount of 
mass. Such a collision is assumed to leave the stellar disk of the primary 
galaxy intact, but induces instabilities in the cold gas present, driving 
it to the centre of the galaxy, where it takes part in a burst. 
Secondly, we assume that star formation induced by mergers produces stars 
with a flat initial mass function (IMF). With a larger proportion of high mass 
stars, the total energy radiated in the ultra-violet per units mass of stars 
produced is increased, thus increasing the amount of radiation heating 
the dust. 
Moreover, the flat IMF produces a higher yield of metals than a 
standard, solar neighbourhood IMF, which means more dust. 

The impact of these two ingredients is readily apparent from the 
comparisons presented in Fig~\ref{fig:s850comp}. One of the beauties 
of semi-analytcal modelling is that certain aspects of the model 
can be switched on and off in order to assess their impact on the 
predictions. These comparisons show that the assumption of a flat 
IMF in starbursts is the main factor responsible for the model 
reproducing the observed counts. The model predictions for the 
redshift distribution of sub-mm sources are shown in 
Fig.~\ref{fig:dndz}. At bright fluxes, the predictions are in 
good agreement with the median redshift determined by Chapman et~al. (2003).

\section{Conclusions}
\label{sec:conc}

The assumption of a flat IMF in starbursts is undoubtedly controversial.
It is therefore important to explore the predictions of the model in 
detail, to find other evidence in support of this choice. The successes of 
the our new model include: 

\begin{itemize} 

\item The reproduction of the properties of the local galaxy population, 
such as the optical and near infrared luminosity functions and the 
distribution 
of disk scalelengths. This is the first hurdle that any realistic model of 
galaxy formation should overcome. Not only does this undermine any claims of 
chicanery when changing model parameters, it also permits a meaningful 
discussion of the descendants of high redshift galaxies. 

\item The recovery of the luminosity function of Lyman break galaxies 
at $z=3$ and $z=4$, with a realistic degree of dust extinction in the 
rest frame UV, computed by tracking the chemical evolution of the model 
galaxies and calculating the sizes of the disk and bulge components. 

\item Nagashima et~al. (2005a) show that the model with a flat IMF 
reproduces the observed abundances of elements in the hot gas in clusters. 

\item Nagashima et~al. (2005b) applied the same model to the calculation 
of element abundances in elliptical galaxies and again found better 
agreement with the model in which starbursts have a flat IMF. 

\item Le Delliou et~al. (2005, 2006) computed the abundance of Lyman-alpha 
emitters using {\tt GALFORM}. The Baugh et~al. model gives a somewhat better 
match to the shape of the observed counts than a model with a standard IMF.  

\end{itemize}

Granato et~al. (2004) present an alternative model in which they consider 
the evolution of quasars and spheroids. These authors find that they can 
explain the number counts of sub-mm galaxies without using a non-standard 
IMF, by using different feedback and gas cooling prescriptions from those 
employed in the model of Baugh et~al. (2005). While it is not clear that these 
recipes would still work in a fully fledged semi-analytical model (Granato et~al. do 
not follow galactic disks nor do they consider mergers between galaxies or 
between haloes),  it will be interesting to see if the new generation of 
semi-analytical models with modified cooling and feedback prescriptions in 
massive haloes can reproduce the number counts of dusty galaxies with a 
standard choice of IMF (Croton et~al. 2006; Bower et~al. 2006).

\acknowledgements  CMB and AJB are supported by the Royal Society; this 
research was also supported by {\tt PPARC}.

\end{document}